\documentclass{ws-procs9x6}

\usepackage{booktabs}

\begin{document}

\title{Vector computation}

\author{Karl Svozil}

\address{Institute for Theoretical Physics, TU Wien,\\
Wiedner Hauptstrasse 8-10/136, 1040 Vienna,  Austria\\
E-mail: svozil@tuwien.ac.at\\
http://tph.tuwien.ac.at/\string~svozil}

\begin{abstract}
Quantum physical resources are directional quantities that can be formalized by unit vectors or the associated orthogonal projection operators. When compared to classical computational states which are elements of (power) sets vector computations offer (dis)advantages.
\end{abstract}


\bodymatter

\section{Epistemology versus ontology in the quantum computation context}

In order to claim practical relevance,
any notion and quantitative means of ``computation'' has to be ultimately grounded in
physics, because \emph{``information is physical''}~\cite{landauer},
and so is the manipulation of information.
The Church-Turing thesis --
in Turing's own words \emph{``a man
provided with paper, pencil, and rubber, and subject to strict discipline, is in
effect a universal machine.''}~\cite{copeland04turing}
--
is a conjecture attempting to achieve just this goal: connecting physics to an appropriate formalism.

Yet such conceptualizations bear, in their very success, a dangerous tendency
to forget about the analogy and instead go for the formalism.
Thereby nature is confused with formalism -- indeed, our own partly formalized narratives about nature
-- just as theater~\cite{Arthaud-en} is taken for life,
or propaganda~\cite{bernais-propaganda} for fact.
This results in beliefs in theoretical and hypothetical conceptual entities,
\emph{``the existence of various religious stigmas and social pressures, that taken
together, amount[[ed]] to an evangelical crusade''}~\cite{clauser-talkvie}.

Thereby, issues related to the formal representability of physical entities and processes
are far from settled, and \emph{``bordering on the mysterious''}~\cite{wigner}.
Yet, unlike Wigner, I believe that there may be at least two handy but mutually contradicting
reasons for the effectiveness of mathematics in the natural sciences: one postulates
that laws ``emerge'' from disorder~\cite{Exner-1908,Yanofsky-chaos,svozil-2018-was}.
Another, apparently converse, reason for lawfulness is that we are inhabiting a virtual reality
simulated by a computational process~\cite{zuse-70,toffoli:79,simula,permutationcity,Bostrom-sim}.
In this hypothesis whatever
``laws of nature'' science might discover should be perceived as epistemic,
intrinsic reflections of, or correspondences to, this ontological, extrinsic (to us) computation.

The above rant served the purpose to suggest, and prepare the Reader for, a cautious disengagement between
epistemology and ontology -- what
might be claimed and believed to be known on the one hand,
and what may be ``lurking behind the detector clicks'' on the other hand.
Even with this proviso, one has to keep in mind that there is no ``Archimedean point''
or ``ontological anchor'' upon which an ``objective reality''
(whatever that is) can be based.

In particular, whenever claims are issued about quantum resources, assets, or features
-- such as quantum parallelism by coherent superposition of classically distinct states,
or entanglement as the relational encoding of multi-partite states --
which might go beyond operationally established classical means and could give rise to quantum advantages,
caution is advisable.
Because in such cases the metaphor might supervene or ``outperform'' the simulated,
yielding improper expectations and overstated, almost evangelical, claims~\cite{svozil-2016-quantum-hokus-pokus}.

\section{Types of quantum oracles for randomness: pure states in a superposition versus mixed states}

\label{2020-vectorc-irrev}

There is a sombre fact of contemporary quantum physics:
as of today the ``measurement problem'', as contemplated by von Neumann~\cite{v-neumann-49},
Schr\"odinger~\cite{schrodinger,schroedinger-interpretation},
and repeated by Everett~\cite{everett,everett-1956} and Wigner~\cite{wigner:mb},
despite numerous attempts to resolve it~\cite{london-Bauer-1983},
remains disputed and unsolved.
One of the issues is the simple mathematical impossibility to achieve irreversibility
-- the notorious ``collapse of the wave function'', or state reduction at measurement point~\cite{mermin-07} --
from a completely reversible temporal evolution.
Indeed this should be indisputable from ``group theory~101'': the concatenation of unitary operators never yields outside of the realm of unitary and thus reversible transformations.
For finite groups, Cayley's theorem states that one essentially is dealing with permutations,
with one-to-one transformations, with re-expressions, re-samplings of the same ``message''.

Stated differently a nesting argument~\cite{v-neumann-49,everett,everett-1956,wigner:mb} essentially ``enlarges'' the domain of reversibility to
include whatever resources or regions of (Hilbert or configuration) space are necessary to re-establish reversibility,
thereby disputing any irreversibility postulated
by quantum mechanics~\cite{PhysRevD.22.879,PhysRevA.25.2208,greenberger2,Nature351,Zajonc-91,PhysRevA.45.7729,PhysRevLett.73.1223,PhysRevLett.75.3783,hkwz}.
For all practical purposes~\cite{bell-a} quantum systems remain ``epistemically'' irreversible~\cite{engrt-sg-I,engrt-sg-II},
but so are classically reversible statistical systems, for which the entropy increase dissolves into thin air
as one looks ``closer'' at individual constituents~\cite{Myrvold2011237}.

Having just avoided the quantum Scylla of ``irreversibility through reversibility''
brings us closer to the quantum Charybdis of ``quantum jellification''
by the prevailing quantum superposition of classically distinct states of matter and mind,
described so vividly in the late Schr\"odinger's Dublin seminars~\cite{schroedinger-interpretation}, repeating the cat paradox~\cite{schrodinger}
in terms of jellyfish.
Without measurement -- how can we and everything around us ``remain stable'' and \emph{not} dissolve into the chasm of coherent superposition,
and how come that, despite this, we experience a fairly unique cognition and presence?

Dirac's ``why bother?'' objection~\cite{bell-a} to all of this might be to not bother at the moment and
leave the task of solving these issues to future generations. Feynman went a step further
and demanded to cease thinking about them, thereby taking the formalism for granted (like a gospel) as given,
and thus effectively shut-up and calculate~\cite{feynman-law,mermin-1989-shutup,mermin-2004-shutup}.

However, such a repression strategy has consequences:
first of all, as mentioned earlier the deterministic, unitary evolution of quantum states (through nesting) contradicts assumptions about irreversible measurements.
Thereby the ontology of the alleged irreducible randomness of single events through measurements of a coherent superposition of classically
distinct or even mutually exclusive states remains open.
Under such circumstances no justification for their stochastic character, no ``quantum certification''~\cite{zeil-05_nature_ofQuantum}
of, in theologic terminology, \emph{creatio continua},
can be given, because consistency issues are unresolved and currently abolished, thereby either relegating a resolution of the argument to the future
or not addressing the ontology at all.
Quantum uncertainty and random outcomes via irreversible measurements of coherent superposition therefore remains conjectural.

Another related ontological issue regards the existence of mixed states.
First of all, the same issues as for measurements are pertinent: how does one obtain a mixed state from pure ones by
one-to-one unitary state evolution?
Claiming to be able to obtain mixed states from pure ones amounts to pretending to get along with outright mathematical incorrectness.

One formal way of generating mixed states from pure multi-partite states
is by taking the partial trace with respect to the Hilbert space of one particle,
a ``beam dump'' of sorts\cite{engrt-sg-I,engrt-sg-II}.
As the trace is essentially a many-to-one operation, irreversibility ensues.
This can be easily corroborated by the non-uniqueness of purification which can be envisioned as the ``quasi-inverse''
(because, strictly speaking, due to non-uniqueness there is no inverse) of the partial trace~\cite[Section~8.3.1]{nielsen-book10}.

So what is the ontologic status of mixed quantum states? This again is unknown and may again,
by the  ``why bother?'' objection~\cite{bell-a},
be relegated until some later times.
Again the question arises: how can we trust quantum random sequences originating from mixed state measurements?

Even more convoluted is the question if there is any criterion or difference between sequences generated
by measurement of pure states which are in a coherent superposition on the one hand,
and by measurement of mixed quantum states on the other hand.
As we cannot even formally grasp consistently the notions of irreversible measurement and production of mixed states
we ought to accept our total loss to conceptualize the differences or similarities between them.
We are left with the hope that, as both issues appear to have the same group-theoretic roots,
any solution of one will, by reduction, entail a solution of the other.

Another ``fashionable'' attempt to ascertain and thereby certify quantum randomness involving ``quantum contextuality''~\cite{svozil-2009-howto,svozil-2017-b}
would be to assume the co-existence of complementary observables
and, relative to the respective implicit assumptions (most notably context independence), prove (by contradiction or statistical demonstration) the impossibility
for any value definiteness of complementary, incompatible observables prior to measurements~\cite{kochen1,froissart-81,pitowsky:218,2015-AnalyticKS,svozil-2017-b}.
Suffice it to say that, as these hypothetical arguments go they are contingent on the respective
counterfactual configurations imagined, and thus appear subjective and inconclusive~\cite{svozil-2020-c}.

\section{Questionable parallelism by views on a vector}

From now on only pure states, represented by (unit) vectors spanning a one-dimensional subspace of a vector space, will be considered.
Schr\"odinger's aforementioned question regarding quantum jellification~\cite{schroedinger-interpretation},
a variant of his earlier ``cat paradox''~\cite{schrodinger}, might be brought to  practical use for quantum parallelization:
for the sake of irritation suppose, as is often alleged in quantum computations, that the many mutually exclusive (classical) states in a coherent superposition
\emph{``be not alternatives but all really happen simultaneously $\ldots$
if the laws of nature took this form for, let
me say, a quarter of an hour, we should find our surroundings rapidly
turning into a quagmire, or sort of a featureless jelly or plasma, all contours
becoming blurred, we ourselves probably becoming jelly fish.''}

One of the simplest answers that can be given to these concerns is that this is just an epistemic question arising from a ``wrong'' viewpoint or perspective.
Because if one chooses an orthonormal basis of Hilbert space of which the state vector is an element then
the coherent superposition reduces to a single, unique term -- namely that unit vector.
Unitary quantum evolution amounts to ``rotating'' this vector in Hilbert space.
There is no ``jellification'' of this state with respect to the ``proper'' bases (for dimensions higher than 2 there is a continuum of bases)
of which the state is one part.
In this view, there is no ``simultaneous existence'' of two or more classically distinct states.
All other probabilistic views are mere (continuity of) multiplicities, projections of sorts.
For a similar, more formalized, vision see Gleason's remarks on taking
\emph{``the square of the norm of the projection''} in the second paragraph of Ref.~\cite{Gleason}.

Indeed, one can find some hints on this solution in Schr\"odinger's own writings on {\it the Vedantic vision}~\cite[chapter~V]{book:1170675}:
\emph{``the plurality that we perceive is only \emph{an appearance; it is not real.}
Vedantic philosophy, in which this is a
fundamental dogma has sought to clarify it by a number of
analogies, one of the most attractive being the many-faceted crystal which, while showing hundreds of little
pictures of what is in reality a single existent object, does
not really multiply that object.''
}

This might be considered bad news for quantum parallelism.
Because if any such parallelism is based upon epistemology
-- about appearance without any substantial reality --
all that remains as a resource is our ability to measure properties of the vector
from a ``different angle'' than the one this vector has been defined.

This is particularly pertinent for the ``extraction'' of information in a coherent superposition
by classical irreversible measurements
(cf. my earlier comments in Section~\ref{2020-vectorc-irrev})
``reducing'' a coherent superposition of a exponential variety of hypothetically conceivable classical distinct states
to a \emph{single} such state.
Because what good is it to contemplate about such a counterfactual variety if we have no direct access to it?

It gets even more problematic as we have postulated that the ``extraction'',
the outcome corresponding to this single state, occurs without sufficient reason~\cite{sep-sufficient-reason} and
thereby eventualizes in an irreducibly stochastic~\cite{zeil-05_nature_ofQuantum} manner,
a situation denoted in theology by {\it creatio continua}:
one can get only a single click or outcome per experiment, corresponding to an ``exponential reduction'' of computational states with respect to the state vector representation
prior to extraction.

\section{Computation by projective measurements of partitioned state space}

Nevertheless, even in this reduced scheme of parallelism,
it might be possible to formulate relational queries corresponding to useful information
by appropriate partitioning of the state space~\cite{DonSvo01,svozil-2002-statepart-prl,svozil-2003-garda}.
We may formulate the fundamental problem of intrinsically operational vector encoding of a computation aka quantum computation:
\emph{under what circumstances is it possible to derive ``useful'' information about (the components) of a vector?}

It is not too unreasonable to suspect that an answer to this question can be given in terms
of \emph{relational properties} of the vector encoding~\cite{2007-tkadlec-svozil-springer,svozil-2019-s}.
Deutsch's algorithm~\cite[Section~1.4.3]{nielsen-book10} as well as the quantum Fourier transform
based on period finding~\cite[Section~5.4.1]{nielsen-book10} may be examples for such relational encodings
realizable by state mismatches (``prepare one state, measure another'').
More generally, the partitioning of finite groups by cosets, in particular, the hidden subgroup problem~\cite[Section~5.4.3]{nielsen-book10}
maybe a way to systematically exploit views on vectors,
but so far there exists only anecdotal evidence~\cite[Figure 5.5., p.~241]{nielsen-book10} for that.

There is a way to extract information from a quantum state by constructing proper (with respect to the computational task or query) subspaces
and the orthogonal projection operators onto such subspaces.
How ought this computational method be understood?
It will be argued that it could be perceived in terms of (equi)partitioning the quantum state space,
and, as mentioned earlier,
by using the respective projection operators as filters~\cite{DonSvo01,svozil-2002-statepart-prl,svozil-2003-garda}.

Suppose it is possible to encode (the solution to) a problem into a Hilbert space spanned by a
collection of pure state vectors encoding functional instances and functional properties into one- and higher-dimensional subspaces thereof.
For instance, a binary function of several bits can be encoded into quantum states
(i) by state vectors whose first components with respect to an orthonormal basis (normalization aside) are either 0 or 1,
depending on whether or not the function on the inputs evaluates to 0 or 1, and
(ii)
later ``auxiliary'' components are added to ensure mutual orthogonality of the state vectors.
That is, in order to obtain an orthonormal basis one could, for instance, employ dimensional lifting, and thereby enlarge the Hilbert space.
This results in an orthonormal (after normalization) basis of the aforementioned Hilbert space.
Therefore the elements of the orthonormal basis represent the individual instances of the function or problem.

Now if one forms the orthogonal projection operator
as the sum of the dyadic products of the respective vectors of the orthonormal basis of the subspace encoding a particular problem or a query,
then this projection operator is capable of solving the property or problem it was encoded to solve in a single run.
All that needs to be done is apply this projection ``filter'' to a state encoding some arbitrary problem instance.

Let me demostrate this by an example which is a generalized Deutsch algorithm.
Consider arbitrary binary functions of $n$ classical bits.
Suppose an unknown arbitrary such function is given, and suppose that the question is not which function exactly it is,
but about a relational property which for instance refers to ``common'' or ``different'' properties of functions of this class; say parity.

How could one find such a particular property without having to identify the respective function completely?
It is not too difficult to argue that there are $2^n$ possible arguments and $2^{2^n}$ such binary functions of $n$ bits.
For the sake of a reasonable ``small'' demonstration, take $n=2$
($n=1$ amounts to Deutsch's problem; cf. Refs.~\cite[Section~2.2]{mermin-07} and~\cite[Section~1.4.3]{nielsen-book10}).
Table~\ref{2020-vectorcomputation-table1} enumerates all binary functions of two classical bits.
(At this point we are not dealing with questions of enlarging the Hilbert space to obtain overall reversibility in case the functions are not reversible.)

In the next step, a system of vectors $\vert {\bf e}_i \rangle$
is obtained by identifying the valuations of the functions on the respective bit values with entries in coordinate tuples,
as enumerated in the second to last column of Table~\ref{2020-vectorcomputation-table1}.
Based on these vectors an orthonormal basis of a (subspace) of a high-dimensional Hilbert space can
be effectively generated by dimensional lifting~\cite{svozil-2016-ggs}, so that
$\vert {\bf e}_i    \rangle \mapsto \vert {\bf b}_i    \rangle$ with $\langle  {\bf b}_i    \vert {\bf b}_j    \rangle =\delta_{ij}$ and $i,j=1,\ldots ,16$
as enumerated (without normalization) in the last column of Table~\ref{2020-vectorcomputation-table1}.
Thereby the zero vector $\vert {\bf e}_1    \rangle$ can, for instance, be ad hoc mapped into a
subspace of this larger dimensional Hilbert space which is orthogonal to all other subspaces
(this may be achieved by adding another dimension).
Note that
\begin{itemize}
\item[(i)] in general dimensional lifting is neither unique --
there exit other, rather inefficient methods~\cite{svozil-2016-vector} (with respect to the number of auxiliary extra dimensions)
to orthogonalize the vectors corresponding to the functions $f_i$.
\item[(ii)] Dimensional lifting does not correspond to a unitary transformation as it intentionally changes the inner products (to become zero) in transit to higher dimensions.
Therefore, if one attempts to encode this kind of problems into orthogonal bases of subspaces of higher dimensional Hilbert spaces
one needs to take care of orthogonality from the very beginning.
That is, there has to be a physically feasible way to map the functions $f_i$ into $\vert {\bf b}_i \rangle$.
\item[(iii)]
Accordingly, any way to map the 16 functions $f_i$ into any kind of system of orthogonal vectors suffices for this method as long as it is physically feasible.
\end{itemize}

In the final step a filter is designed which models the binary question by projecting the answers onto the appropriate subspace of the (sub)space spanned by the orthonormal
basis $\big\{ \vert {\bf b}_1 \rangle , \ldots ,  \vert  {\bf b}_{16}   \rangle   \big\}$
such that the question can be answered in a single query.

\begin{sidewaystable}
\vspace*{+250pt}
\tbl{The 16 binary functions of two classical bits. Dots as vector components represent ``very large'' numbers.\label{2020-vectorcomputation-table1}}
{\normalsize
%\resizebox{\textwidth}{!}{
\begin{tabular}{ccccccc}
\toprule
$f_\#$ &00&01&10&11 & corresponding vector & vector after dimensional lifting (20 dimensions) \\
\midrule
$f_1   $&0  &  0 &   0 &   0  & $\vert {\bf e}_1    \rangle=\begin{pmatrix} 0  ,  0 ,   0 ,   0  \end{pmatrix}^\intercal$&{\scriptsize$\vert {\bf b}_1    \rangle=\begin{pmatrix}0 , 0 , 0 , 0 , 1 , 0 , 0 , 0 , 0 , 0 , 0 , 0 , 0 , 0 , 0 , 0 , 0 , 0 , 0 , 0 \end{pmatrix}^\intercal                                           $}\\
$f_2   $&0  &  0 &   0 &   1  & $\vert {\bf e}_2    \rangle=\begin{pmatrix} 0  ,  0 ,   0 ,   1  \end{pmatrix}^\intercal$&{\scriptsize$\vert {\bf b}_2    \rangle=\begin{pmatrix}0 , 0 , 0 , 1 , 0 , 1 , 0 , 0 , 0 , 0 , 0 , 0 , 0 , 0 , 0 , 0 , 0 , 0 , 0 , 0 \end{pmatrix}^\intercal                                           $}\\
$f_3   $&0  &  0 &   1 &   0  & $\vert {\bf e}_3    \rangle=\begin{pmatrix} 0  ,  0 ,   1 ,   0  \end{pmatrix}^\intercal$&{\scriptsize$\vert {\bf b}_3    \rangle=\begin{pmatrix}0 , 0 , 1 , 0 , 0 , 0 , 1 , 0 , 0 , 0 , 0 , 0 , 0 , 0 , 0 , 0 , 0 , 0 , 0 , 0 \end{pmatrix}^\intercal                                           $}\\
$f_4   $&0  &  0 &   1 &   1  & $\vert {\bf e}_4    \rangle=\begin{pmatrix} 0  ,  0 ,   1 ,   1  \end{pmatrix}^\intercal$&{\scriptsize$\vert {\bf b}_4    \rangle=\begin{pmatrix}0 , 0 , 1 , 1 , 0 , -1 , -1 , 1 , 0 , 0 , 0 , 0 , 0 , 0 , 0 , 0 , 0 , 0 , 0 , 0 \end{pmatrix}^\intercal                                         $}\\
$f_5   $&0  &  1 &   0 &   0  & $\vert {\bf e}_5    \rangle=\begin{pmatrix} 0  ,  1 ,   0 ,   0  \end{pmatrix}^\intercal$&{\scriptsize$\vert {\bf b}_5    \rangle=\begin{pmatrix}0 , 1 , 0 , 0 , 0 , 0 , 0 , 0 , 1 , 0 , 0 , 0 , 0 , 0 , 0 , 0 , 0 , 0 , 0 , 0 \end{pmatrix}^\intercal                                           $}\\
$f_6   $&0  &  1 &   0 &   1  & $\vert {\bf e}_6    \rangle=\begin{pmatrix} 0  ,  1 ,   0 ,   1  \end{pmatrix}^\intercal$&{\scriptsize$\vert {\bf b}_6    \rangle=\begin{pmatrix}0 , 1 , 0 , 1 , 0 , -1 , 0 , -2 , -1 , 1 , 0 , 0 , 0 , 0 , 0 , 0 , 0 , 0 , 0 , 0 \end{pmatrix}^\intercal                                        $}\\
$f_7   $&0  &  1 &   1 &   0  & $\vert {\bf e}_7    \rangle=\begin{pmatrix} 0  ,  1 ,   1 ,   0  \end{pmatrix}^\intercal$&{\scriptsize$\vert {\bf b}_7    \rangle=\begin{pmatrix}0 , 1 , 1 , 0 , 0 , 0 , -1 , -2 , -1 , -6 , 1 , 0 , 0 , 0 , 0 , 0 , 0 , 0 , 0 , 0 \end{pmatrix}^\intercal                                       $}\\
$f_8   $&0  &  1 &   1 &   1  & $\vert {\bf e}_8    \rangle=\begin{pmatrix} 0  ,  1 ,   1 ,   1  \end{pmatrix}^\intercal$&{\scriptsize$\vert {\bf b}_8    \rangle=\begin{pmatrix}0 , 1 , 1 , 1 , 0 , -1 , -1 , -4 , -1 , -12 , -84 , 1 , 0 , 0 , 0 , 0 , 0 , 0 , 0 , 0 \end{pmatrix}^\intercal                                   $}\\
$f_9   $&1  &  0 &   0 &   0  & $\vert {\bf e}_9    \rangle=\begin{pmatrix} 1  ,  0 ,   0 ,   0  \end{pmatrix}^\intercal$&{\scriptsize$\vert {\bf b}_9    \rangle=\begin{pmatrix}1 , 0 , 0 , 0 , 0 , 0 , 0 , 0 , 0 , 0 , 0 , 0 , 1 , 0 , 0 , 0 , 0 , 0 , 0 , 0 \end{pmatrix}^\intercal                                           $}\\
$f_{10}$&1  &  0 &   0 &   1  & $\vert {\bf e}_{10} \rangle=\begin{pmatrix} 1  ,  0 ,   0 ,   1  \end{pmatrix}^\intercal$&{\scriptsize$\vert {\bf b}_{10} \rangle=\begin{pmatrix}1 , 0 , 0 , 1 , 0 , -1 , 0 , -2 , 0 , -6 , -40 , -3442 , -1 , 1 , 0 , 0 , 0 , 0 , 0 , 0 \end{pmatrix}^\intercal                                 $}\\
$f_{11}$&1  &  0 &   1 &   0  & $\vert {\bf e}_{11} \rangle=\begin{pmatrix} 1  ,  0 ,   1 ,   0  \end{pmatrix}^\intercal$&{\scriptsize$\vert {\bf b}_{11} \rangle=\begin{pmatrix}1 , 0 , 1 , 0 , 0 , 0 , -1 , -2 , 0 , -4 , -30 , -2578 , -1 , -8874706 , 1 , 0 , 0 , 0 , 0 , 0 \end{pmatrix}^\intercal                          $}\\
$f_{12}$&1  &  0 &   1 &   1  & $\vert {\bf e}_{12} \rangle=\begin{pmatrix} 1  ,  0 ,   1 ,   1  \end{pmatrix}^\intercal$&{\scriptsize$\vert {\bf b}_{12} \rangle=\begin{pmatrix}1 , 0 , 1 , 1 , 0 , -1 , -1 , -4 , 0 , -10 , -70 , -6020 , -1 , -20723712 , \cdot , 1 , 0 , 0 , 0 , 0 \end{pmatrix}^\intercal                   $}\\
$f_{13}$&1  &  1 &   0 &   0  & $\vert {\bf e}_{13} \rangle=\begin{pmatrix} 1  ,  1 ,   0 ,   0  \end{pmatrix}^\intercal$&{\scriptsize$\vert {\bf b}_{13} \rangle=\begin{pmatrix}1 , 1 , 0 , 0 , 0 , 0 , 0 , 0 , -1 , -2 , -14 , -1202 , -1 , -4137858 , \cdot , \cdot , 1 , 0 , 0 , 0 \end{pmatrix}^\intercal                   $}\\
$f_{14}$&1  &  1 &   0 &   1  & $\vert {\bf e}_{14} \rangle=\begin{pmatrix} 1  ,  1 ,   0 ,   1  \end{pmatrix}^\intercal$&{\scriptsize$\vert {\bf b}_{14} \rangle=\begin{pmatrix}1 , 1 , 0 , 1 , 0 , -1 , 0 , -2 , -1 , -8 , -54 , -4644 , -1 , -15986864 , \cdot , \cdot , \cdot , 1 , 0 , 0 \end{pmatrix}^\intercal            $}\\
$f_{15}$&1  &  1 &   1 &   0  & $\vert {\bf e}_{15} \rangle=\begin{pmatrix} 1  ,  1 ,   1 ,   0  \end{pmatrix}^\intercal$&{\scriptsize$\vert {\bf b}_{15} \rangle=\begin{pmatrix}1 , 1 , 1 , 0 , 0 , 0 , -1 , -2 , -1 , -6 , -44 , -3780 , -1 , -13012562 , \cdot ,\cdot , \cdot , \cdot , 1 , 0 \end{pmatrix}^\intercal         $}\\
$f_{16}$&1  &  1 &   1 &   1  & $\vert {\bf e}_{16} \rangle=\begin{pmatrix} 1  ,  1 ,   1 ,   1  \end{pmatrix}^\intercal$&{\scriptsize$\vert {\bf b}_{16} \rangle=\begin{pmatrix}1 , 1 , 1 , 1 , 0 , -1 , -1 , -4 , -1 , -12 , -84 , -7222 , -1 , -24861568 , \cdot , \cdot , \cdot , \cdot , \cdot , 1 \end{pmatrix}^\intercal  $}\\
\bottomrule
\end{tabular} }
% }
%\vspace*{+13pt}
\end{sidewaystable}

Suppose the question is to find the parity of a function $f_i(x,y) \in \{0,1\})$ with $x,y \in \{0,1\}$, $i\in \{1,\ldots ,16\}$.
All we need to do is to partition the functional space
$\big\{   f_1  , \ldots ,   f_{16}   \big\}$
into functions with an even or an odd number of outputs ``1''. More explicitly, for the functions enumerated in Table~\ref{2020-vectorcomputation-table1} and for parity,
the partition is
\begin{equation}
 \Big\{
\big\{
 f_2, f_3,f_5,f_8,f_9,f_{12},f_{14},f_{15}
\big\}
,
\big\{
 f_1, f_4,f_6,f_7,f_{10},f_{11},f_{13},f_{16}
\big\}
\Big\}
,
\end{equation}
corresponding to the orthogonal projection operators
\begin{equation}
\begin{split}
\textsf{\textbf{E}}_1 =
 \vert \mathbf{b}_2 \rangle \langle \mathbf{b}_2 \vert
+\vert \mathbf{b}_3 \rangle \langle \mathbf{b}_3 \vert
+\vert \mathbf{b}_5 \rangle \langle \mathbf{b}_5 \vert
+\vert \mathbf{b}_8 \rangle \langle \mathbf{b}_8 \vert \qquad \qquad \\
+\vert \mathbf{b}_9 \rangle \langle \mathbf{b}_9 \vert
+\vert \mathbf{b}_{12} \rangle \langle \mathbf{b}_{12} \vert
+\vert \mathbf{b}_{14} \rangle \langle \mathbf{b}_{14} \vert
+\vert \mathbf{b}_{15} \rangle \langle \mathbf{b}_{15} \vert
,\text{ and }\\
\textsf{\textbf{E}}_0 =  \textsf{\textbf{1}} - \textsf{\textbf{E}}_1 =
 \vert \mathbf{b}_1 \rangle \langle \mathbf{b}_1 \vert
+\vert \mathbf{b}_4 \rangle \langle \mathbf{b}_4 \vert
+\vert \mathbf{b}_6 \rangle \langle \mathbf{b}_6 \vert
+\vert \mathbf{b}_7 \rangle \langle \mathbf{b}_7 \vert \qquad \qquad \\
+\vert \mathbf{b}_{10} \rangle \langle \mathbf{b}_{10} \vert
+\vert \mathbf{b}_{11} \rangle \langle \mathbf{b}_{11} \vert
+\vert \mathbf{b}_{13} \rangle \langle \mathbf{b}_{13} \vert
+\vert \mathbf{b}_{16} \rangle \langle \mathbf{b}_{16} \vert
.
\end{split}
\end{equation}
The parity of an unknown given binary function of two bits
can be obtained by a single query measuring the
propositional ``parity property''
associtated with the observable $\textsf{\textbf{E}}_1  = \textsf{\textbf{1}} - \textsf{\textbf{E}}_0$.
In principle this method can be generalized to the parity problem of binary functions of $n$ bits,
utilizing a parallelization of the order of $2^{2^n}$ at the cost of expanding Hilbert space to about twice this number of dimensions.
It remains to be seen whether this method violates the assumptions in Ref.~\cite{Farhi-98}.
In any case it should be noted that parity is an example of a much wider problem class associated with relational properties
which can be represented or parametrized by partitioning appropriate subspaces of Hilbert space.

\section{Entanglement as relational parallelism across multi-partite states}

From a purely formal point of view, entangled particles are modeled by the
indecomposability of state vectors in a Hilbert space which is a non-trivial
tensor product of two or more Hilbert spaces.
Indecomposability means that the respective state vector cannot be decomposed into
a single product of factors of the states of the constituent particles.
Instead, an entangled state can be written as the coherent superposition (aka linear combination) of such product states.

This immediately suggests that whenever such indecomposable vectors occur they can be ``rotated'' by unitary transformations
into a single product form; say a vector of the Cartesian standard basis in the respective tensor product of two or more Hilbert spaces.
Any such transformation cannot be expected to be acting ``locally'' in a single constituent space but rather ``globally''
across the single constituent spaces.

Physically this means that we are not dealing with single-particle properties but again with \emph{relational properties};
whereby relational information is encoded across the multiple constituents of such a state~\cite{schrodinger,zeil-99}.
This can be expected: as unitary transformations are defined by
rotations transforming some orthonormal basis into another orthonormal basis~\cite{Schwinger.60}
this amounts to a kind of ``zero-sum game'' between \emph{localized} information aka properties on \emph{individual} constituents
on the one hand, and  \emph{relational} information which for instance refers to ``common'' or ``different'' properties within
\emph{collectives or groups} of constituents on the other hand.

\section{On partial views of vectors}

In the context of state purification the following more general question arises:
What kind of value might a partial knowledge of or about a vector have?
After all, embedded observers~\cite{toffoli:79} may obtain only partial knowledge and control
of the degrees of freedom entailed by overseeing a ``small'' subset of a ``much larger'' Hilbert space they inhabit.

Suppose an observer has acquired knowledge about an incomplete list of components (relative to a particular basis) of a pure state vector.
This can be formalized either by a projection of this vector onto a subspace of the Hilbert space
or by ``extraction'' of the coordinates by the respective vectors of the dual basis of the dual space.

One possibility would be to ``complete'' the state vector by various procedures,
such as the aforementioned purification.
Alternatively one might consider the subspace spanned by both the given and the missing basis elements~\cite{havlicek-svozil-2020-dec}.

\section{Summary}

I have presented a revisionist glance at quantum computation as seen from an equally revisionist perspective on quantum theory.
The main  t\'elos, that is, the end, purpose, or goal, of these considerations rests in the emphasis that only with a proper understanding
of the quantum physical resources it is possible to develop a comprehensive theory of quantum information and computation.

For instance, the mere communal or individual canonical believe in ontological quantum randomness -- without mentioning the
implicit assumptions or corroborations which are essentially based upon incapacity; that is the experience that nobody so far has come up with any causes and
necessary and sufficient reasons for quantum outcomes -- suggests that any such claims need be viewed as epistemic, anecdotal and preliminary.

I believe that quantum information and computation are intimately tied to foundational issues.
Therefore, such issues need at least to be kept in mind if one assesses the capacities of quantum systems to store and process information.

%\begin{acknowledgments}
\section*{Acknowledgments}
Parts of this discussion have been inspired by a discussion with Noson Yanofsky about
whether quantum superpositions and entanglements make the universe unboring.

The author acknowledges the support by the Austrian Science Fund (FWF): project I 4579-N and the Czech Science Foundation (GA\v CR): project 20-09869L,
%\end{acknowledgments}

%\bibliographystyle{ws-procs9x6}
% \bibliographystyle{revtex}
% \bibliography{svozil}

\end{document}